\documentclass[preprint,showpacs,preprintnumbers,amsmath,amssymb]{revtex4}
\usepackage{booktabs}
\usepackage{mathrsfs}
\usepackage{epsfig}
\usepackage{graphicx}
\usepackage{dcolumn}
\usepackage{bm}
\usepackage{amsmath}
\usepackage{slashed}       

\let\jnfont=\rm
\def\NPB#1,{{\jnfont Nucl.\ Phys.\ B }{\bf #1},}
\def\PLB#1,{{\jnfont Phys.\ Lett.\ B }{\bf #1},}
\def\EPJC#1,{{\jnfont Eur.\ Phys.\ Jour.\ C }{\bf #1},}
\def\PRD#1,{{\jnfont Phys.\ Rev.\ D }{\bf #1},}
\def\PRL#1,{{\jnfont Phys.\ Rev.\ Lett.\ }{\bf #1},}
\def\MPLA#1,{{\jnfont Mod.\ Phys.\ Lett.\ A }{\bf #1},}
\def\JPG#1,{{\jnfont J.\ Phys.\ G}{\bf #1},}
\def\CTP#1,{{\jnfont Commun.\ Theor.\ Phys.\ }{\bf #1},}
\def\ZPC#1,{{\jnfont Z.\ Phys.\ C }{\bf #1},}
\def\JHEP#1,{{\jnfont JHEP \ }{\bf #1},}
\def\lsim{\raise0.3ex\hbox{$<$\kern-0.75em\raise-1.1ex\hbox{$\sim$}}}
\def\gsim{\raise0.3ex\hbox{$>$\kern-0.75em\raise-1.1ex\hbox{$\sim$}}}

\begin{document}

\title{Pair Production of a 125 GeV Higgs Boson in MSSM and NMSSM at the LHC}

\author{Junjie Cao$^{1,2}$, Zhaoxia Heng$^1$, Liangliang Shang$^1$, Peihua Wan$^1$, Jin Min Yang$^3$ }
\affiliation{
  $^1$  Department of Physics,
        Henan Normal University, Xinxiang 453007, China \\
  $^2$ Center for High Energy Physics, Peking University,
       Beijing 100871, China \\
  $^3$ State Key Laboratory of Theoretical Physics,
      Institute of Theoretical Physics, Academia Sinica, Beijing 100190, China
      \vspace{1cm}}

\begin{abstract}
In light of the recent LHC Higgs search data,
we investigate the pair production of a SM-like Higgs boson around 125 GeV
in the MSSM and NMSSM. We first scan the parameter space of each model
by considering various experimental constraints, and then calculate
the Higgs pair production rate in the allowed parameter space.
We find that in most cases
the dominant contribution to the Higgs pair production comes from the
gluon fusion process and the production rate can be greatly enhanced,
maximally 10 times larger than the SM prediction
(even for a TeV-scale stop the production rate can still be enhanced by
a factor of 1.3).
We also calculate the $\chi^2$ value with the current Higgs data
and find that in the most favored parameter region the production
rate is enhanced by a factor of 1.45 in the MSSM,
 while in the NMSSM the production rate can be enhanced or suppressed
($\sigma_{SUSY}/\sigma_{SM}$ varies from 0.7 to 2.4).

\end{abstract}


\maketitle

\section{Introduction}
Based on the combined data collected at the center-of-mass energies of 7 TeV and 8 TeV,
the experimental programme to probe the mechanism of electroweak symmetry breaking
at the LHC has recently witnessed the discovery of a new particle around
125 GeV~\cite{1207ATLAS-CMS}.
The properties of this particle, according to the updated analyses of
the ATLAS and CMS collaborations
at the end of 2012~\cite{1212ATLAS-CMS}, roughly agree with the
the Standard Model (SM) prediction and thus it should play a role in both the
symmetry breaking and the mass generation.
However, the issue of whether this particle is the SM Higgs boson is still open,
and indeed there are some motivations, such as the gauge hierarchy problem
and the excess in the di-photon channel over the SM prediction
\cite{1207ATLAS-CMS,1212ATLAS-CMS}, to consider new physics interpretation
of this boson. Studies in this direction have been performed intensively
in low energy supersymmetry (SUSY) and it was found that some SUSY models
can naturally provide a 125 GeV Higgs boson
\cite{Feb-Cao,Carena-Higgsmass,1213-125GeV-Higgs},
and fit the data better than the SM \cite{July-Cao}
(similar studies have also been performed in some non-SUSY models like the
little Higgs models and two-Higgs-doublet or Higgs-triplet models 
\cite{125-other}).

After the discovery of the Higgs boson, the next important task for the LHC is to test
the property of this Higgs boson by measuring all the possible
production and decay channels with high luminosity.
Among the production channels, the Higgs pair production is a rare process
at the LHC. Since it can play an important role for testing the Higgs self-couplings
~\cite{SM-NLO-35fb,1213-DHiggs}
(the determination of the Higgs self-couplings is of great importance since it is
indispensable to reconstruct the Higgs potential), it will be measured at the LHC with
high luminosity.

In the SM the Higgs pair production at the LHC
proceeds by the parton process $gg \to h h$ through the heavy quark
induced box diagrams and also through the production of an off-shell Higgs which
subsequently splits into two on-shell Higgs bosons \cite{SM-LO-20fb,DHiggsInSM}.
The production rate is rather low for
$\sqrt{s} = 14 {\rm TeV}$, about 20 fb  at leading order~\cite{SM-LO-20fb}
and reaching roughly 35 fb after including the next-to-leading order QCD
correction~\cite{SM-NLO-35fb}. The capability of the LHC to detect this production
process was investigated in~\cite{bbgaga,bbWW,bbtautau,DHiggs-Detect}. These analyses
showed that for a 125 GeV Higgs boson the most efficient channel
is $g g \to  h h \to b \bar{b} \gamma \gamma$ with 6 signal events over
14 background events expected for 600 fb$^{-1}$ integrated luminosity after considering
some elaborate cuts~\cite{bbgaga} (the detection through other channels like
$h h \to b \bar{b} W^+ W^-$ and $h h \to b \bar{b} \tau^+ \tau^-$
has also been studied recently~\cite{bbWW,bbtautau}).
In principle, the capability can be further improved if
the recently developed jet substructure technique~\cite{JetSubstructure} is applied for
the Higgs tagging.

The Higgs pair production at the LHC may also be a sensitive probe for 
new physics.
In supersymmetric models such as the Minimal Supersymmetric Standard Model (MSSM)~\cite{MSSM},
the pair production of the SM-like Higgs boson receives additional contributions from
the loops of
the third generation squarks and also from the parton process $b \bar{b} \to H_i \to h h$
with $H_i$ denoting a CP-even non-standard Higgs boson~\cite{RunningMass,DHiggsInMSSM}.
It was found that in some cases (e.g., a light stop with a large trilinear soft
breaking parameter $A_t$ and/or a large $\tan \beta$ together with moderately light $H_i$),
these new contributions may be far dominant over the SM contribution, and as a result, the
rate of the pair production may be enhanced by several orders \cite{RunningMass,DHiggsInMSSM}.
Note that since the experimental constraints (direct or indirect) on the SUSY parameter space
have been becoming more and more stringent, the previous MSSM results should be updated
by considering the latest constraints. This is one aim of this work.
To be specific, we will consider the following new constraints:
\begin{itemize}
\item The currently measured Higgs boson mass $m_h = 125$ GeV
      \cite{1212ATLAS-CMS}. In SUSY this mass is sensitive to radiative 
      correction and thus the third generation squark
      sector has been tightly limited.
\item The LHC search for the third generation squarks~\cite{1213ThirdSquark-LHC}. So far although
       the relevant bounds are rather weak and usually hypothesis-dependent,
       it becomes more and more clear that a stop lighter than about 200 GeV
       is strongly disfavored.
\item The observation of $B_s \to \mu^+ \mu^-$ by the LHCb~\cite{1213Bsmumu}.
      In the MSSM it is well known that the branching ratio of $B_s \to \mu^+ \mu^-$
       is proportional to $\tan^6\beta/m_H^4$ for a large $\tan \beta$ and a moderately
       light $H$~\cite{Bobeth}. Since the experimental value of $B_s \to \mu^+ \mu^-$
       coincides well with the SM prediction, $\tan \beta$ as a function of $m_{H}$
       has been upper bounded.
\item The LHC search for a non-standard Higgs boson $H$ through the process
      $p p \to H \to \tau^+ \tau^-$~\cite{Htautau}.
      Such the search relies on the enhanced $H \bar{b} b$ coupling and the nought signal
      seen by the LHC
      experiments implies that a broad region in the $\tan \beta-m_H$ plane has been ruled out.
\item The global fit of the SUSY predictions on various Higgs signals to the Higgs data reported
      by the ATLAS and CMS collaborations~\cite{GlobalFit}, the dark matter relic density~\cite{WMAP}
      as well as the XENON2012 dark matter search results~\cite{XENON2012} can also limit SUSY
      parameters in a complex way.  ¡¡
\end{itemize}
Another motivation of this work comes from the fact
that the Next-to-Minimal Supersymmetric Standard Model (NMSSM)~\cite{NMSSM}
is found to be more favored by the Higgs
data and the fine-tuning argument \cite{July-Cao}.
So far the studies on the Higgs pair production
in the NMSSM are still absent. So it is necessary to extend the study to the NMSSM.

This paper is organized as follows. In Sec.~II we briefly introduce the
features of the Higgs sector in the MSSM and NMSSM. Then in Sec.~III
we present our results for the Higgs pair production in both models.
Some intuitive understandings on the results are also presented.
Finally, we summarize our conclusions in Sec.~IV.

\section{Higgs sector in MSSM and NMSSM}
As the most economical realization of SUSY in particle physics,
the MSSM~\cite{MSSM} has been intensively studied. However, since this model
suffers from some problems such as the unnaturalness of $\mu$
parameter, it is well motivated to go beyond this minimal framework.
Among the extensions of the MSSM, the NMSSM as the simplest extension by singlet
field~\cite{NMSSM} has been paid much attention. The
differences between the two models come from their superpotentials
and soft-breaking terms, which are given by
\begin{eqnarray}
 W_{\rm MSSM}&=& Y_u\hat{Q}\cdot\hat{H_u}\hat{U}-Y_d \hat{Q}\cdot\hat{H_d}\hat{D}
-Y_e \hat{L}\cdot\hat{H_d} \hat{E} + \mu \hat{H_u}\cdot \hat{H_d}, \label{MSSM-pot}\\
 W_{\rm NMSSM}&=&Y_u\hat{Q}\cdot\hat{H_u}\hat{U}-Y_d \hat{Q}\cdot\hat{H_d}\hat{D}
-Y_e \hat{L}\cdot\hat{H_d} \hat{E} + \lambda\hat{H_u} \cdot \hat{H_d} \hat{S}
 + \frac{1}{3}\kappa \hat{S^3},\\
 V_{\rm soft}^{\rm MSSM}&=&\tilde m_u^2|H_u|^2 + \tilde m_d^2|H_d|^2
+ (B\mu H_u\cdot H_d + h.c.),\\
V_{\rm soft}^{\rm NMSSM}&=&\tilde m_u^2|H_u|^2 + \tilde m_d^2|H_d|^2
+ \tilde m_S^2|S|^2 +(A_\lambda \lambda SH_u\cdot H_d
+\frac{A_\kappa}{3}\kappa S^3 + h.c.).
\end{eqnarray}
Here $\hat{H}_i$ ($i=u,d$) and $\hat{S}$ denote gauge doublet and
singlet Higgs superfields respectively, $\hat{Q}$, $\hat{U}$, $\hat{D}$, $\hat{L}$ and
$\hat{E}$ represent matter superfields with $Y_i$ ($i=u,d,e$) being their Yukawa
coupling coefficients, $\tilde{m}_i$ ($i=u,d,S$), $B$, $A_\lambda$, and $A_\kappa$ are all
soft-breaking parameters and the dimensionless parameters $\lambda$ and $\kappa$ reflect coupling strengthes of
Higgs self interactions. Note the $\mu$-term in the MSSM is replaced by Higgs
self interactions in the NMSSM, so when the singlet field $\hat{S}$ develops a
vacuum expectation value $s$, an effective $\mu$ is generated by $\mu_{eff} = \lambda s$.

Like the general treatment of the multiple-Higgs theory, one can write the Higgs fields in
the NMSSM as
\begin{eqnarray}
H_u = \left ( \begin{array}{c} H_u^+ \\
       v_u +\frac{ \phi_u + i \varphi_u}{\sqrt{2}}
        \end{array} \right),~~
H_d & =& \left ( \begin{array}{c}
             v_d + \frac{\phi_d + i \varphi_d}{\sqrt{2}}\\
             H_d^- \end{array} \right),~~
S  =  s + \frac{1}{\sqrt{2}} \left(\sigma + i \xi \right),
\end{eqnarray}
and diagonalize their mass matrices to get Higgs mass eigenstates:
\begin{eqnarray} \left( \begin{array}{c} H_1 \\
H_2 \\ H_3 \end{array} \right) = U_H \left( \begin{array}{c} \phi_u
\\ \phi_d\\ \sigma\end{array} \right),~ \left(\begin{array}{c} A_1\\
A_2\\ G^0 \end{array} \right) = U_A \left(\begin{array}{c} \varphi_u
\\ \varphi_d \\ \xi \end{array} \right),~ \left(\begin{array}{c} H^+
\\G^+ \end{array}  \right) =U_C \left(\begin{array}{c}H_u^+\\ H_d^+
\end{array} \right).  \label{rotation}
\end{eqnarray}
Here $H_1$, $H_2$, $H_3$ with convention $m_{H_1}<m_{H_2}<m_{H_3}$
and $A_1$, $A_2$ with convention $m_{A_1} < m_{A_2}$ denote the physical CP-even
and CP-odd Higgs bosons respectively,  $G^0$ and $G^+$ are Goldstone bosons eaten by
$Z$ and $W$ bosons respectively, and $H^+$ is the physical charged Higgs boson.
The Higgs sector in the MSSM can be treated in a similar way except that it
predicts only two physical CP-even states and one physical CP-odd state,
and consequently, the rotation matrices $U_H$ and $U_A$ are reduced to
$2 \times 2$ matrices.

\begin{figure}[thbp]
\includegraphics[width=12cm]{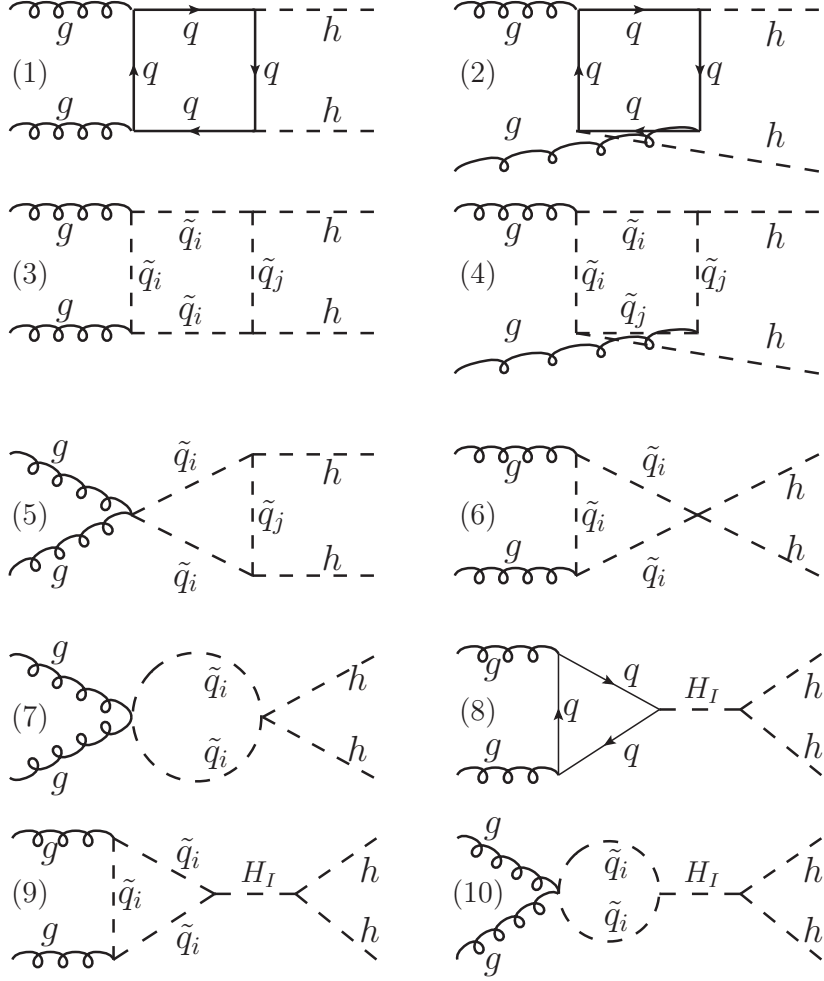}
\vspace{-0.5cm} 
\caption{Feynman diagrams for the pair production of the SM-like Higgs boson
via gluon fusion in the MSSM and NMSSM with $H_I$
denoting a CP-even Higgs ($I=1,2$ for the MSSM and $I=1,2,3$
for the NMSSM) and $\tilde q_{i,j}$ ($i,j=1,2$) for a squark.
The diagrams with initial gluons or final Higgs bosons interchanged
are not shown here. For the quarks and squarks we only consider 
the third generation due to their large Yukawa couplings.}
\label{fig-diagram-total}
\end{figure}

One distinct feature of the MSSM is that $H_1$ usually acts as
the SM-like Higgs boson (denoted by $h$ hereafter) and its mass is upper bounded by $m_Z$ at
tree level. Obviously, to coincide with the LHC discovery of a 125 GeV boson, large
radiative correction to $m_h$ is needed, which in turn usually requires the trilinear soft
breaking parameter $A_t$ to be large. For example, in the
case of large $m_A$ and moderate $\tan\beta$, $m_h$ is given by \cite{Carena-Higgsmass}
\begin{equation}\label{mh}
 m^2_{h}  \simeq M^2_Z\cos^2 2\beta +
  \frac{3m^4_t}{4\pi^2v^2} \left[\ln\frac{m^2_{\tilde t}}{m^2_t} +
\frac{X^2_t}{m^2_{\tilde t}} \left( 1 - \frac{X^2_t}{12m^2_{\tilde t}}\right)\right],
\end{equation}
where the first term is the tree-level mass and the last two terms are the dominant corrections
from the top-stop sector, $m_{\tilde t} = \sqrt{m_{\tilde{t}_1}m_{\tilde{t}_2}}$ ($m_{\tilde{t}_i}$
denotes stop mass with convention $m_{\tilde{t}_1} < m_{\tilde{t}_2}$) represents the
average stop mass scale and $X_t \equiv A_t - \mu \cot\beta$.
One can easily check that for a 500 GeV and 1 TeV stop, 
$|A_t|$ should be respectively about 1.8 TeV and 3.5 TeV
to give $m_h \simeq 125~{\rm GeV}$.

In the NMSSM, $m_h$ exhibits at least two new features \cite{Feb-Cao}. One is that it gets additional
contribution at tree level so that $m_{h,tree}^2 = (m_Z^2 - \lambda^2 v^2 ) \cos^2 2 \beta + \lambda^2 v^2$,
and for $\lambda \sim 0.7 $ and $\tan \beta \sim 1$, $m_h$ can reach 125 GeV even without
the radiative correction. The other feature is that the mixing between the doublet and
singlet Higgs fields can significantly alter the mass. To be more explicit, if the state $H_1$ is $h$,
the mixing is to pull down the mass, while if $H_2$ acts as $h$, the mixing will push up the mass.
Another remarkable character of the NMSSM is that in the limit of very small $\lambda$ and $\kappa$
(but keep $\mu$ fixed),
the singlet field decouples from the theory so that the phenomenology of the
NMSSM reduces to the MSSM. So in order to get a Higgs sector significantly different from
the MSSM, one should consider a large $\lambda$.

Throughout this work, we require $0.50 \leq \lambda \leq 0.7$ in our discussion of the NMSSM
and we consider two scenarios:
\begin{itemize}
\item NMSSM1 scenario: $H_1$ acts as the SM-like Higgs boson. For this
scenario, the additional tree-level contribution to $m_h$ is canceled by the mixing effect,
and if the mixing effect is dominant, the parameters in the stop sector
will be tightly limited in order to give $m_h \simeq 125 {\rm ~GeV}$.
\item NMSSM2 scenario: $H_2$ acts as the SM-like Higgs boson. In this scenario, both
the additional tree-level contribution and the mixing effect can push up the mass.
So for appropriate values of $\lambda$ and
$\tan \beta$, $m_h$ can easily reach 125 GeV even without the radiative correction.
\end{itemize}

\begin{figure}
\includegraphics[width=15cm]{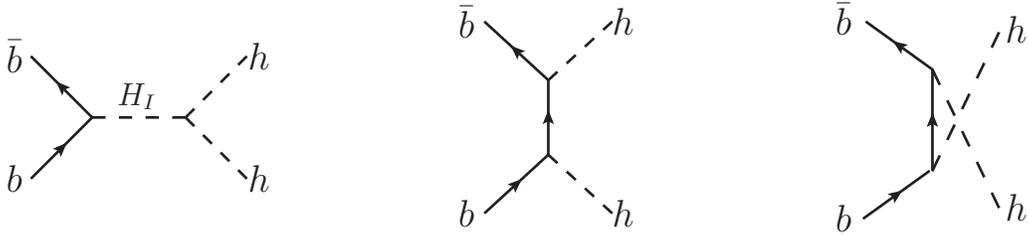}
\vspace{-0.5cm}
\caption{Feynman diagrams for the parton process $b\bar b\to hh$ in the MSSM and NMSSM.}
\label{fig-diagram-bb}
\end{figure}

\section{Calculations and numerical results}
In SUSY the pair production of the SM-like Higgs boson proceeds
through the gluon fusion shown in
Fig.\ref{fig-diagram-total} and the $b\bar b$ annihilation shown in
Fig.\ref{fig-diagram-bb}.
These diagrams indicate that the genuine SUSY contribution to the amplitude
is of the same perturbation order as the SM contribution. So
the SUSY prediction on the production rate may significantly deviate from the SM result.
To ensure the correctness of our calculation,
we checked that we can reproduce the SM results presented in
\cite{SM-LO-20fb} and the MSSM results in \cite{RunningMass}.
Since the analytic expressions are quite lengthy, we do not present
here their explicit forms.

In our numerical calculation
we take $m_t=173$~GeV, $m_b=4.2$~GeV, $m_Z=91.0$~GeV,
$m_W=80.0$~GeV~and~$\alpha=1/128$~\cite{PDG}, and use CT10~\cite{CT10} to
generate the parton distribution functions with the renormalization scale
$\mu_R$ and the factorization scale $\mu_F$ chosen to be $2 m_h$.
The collision energy of the LHC is fixed to be 14 ${\rm TeV}$.
Then we find that for $m_h=125$~GeV,
the production rate in the SM is 18.7 fb for $gg\to hh$
and  0.02 fb for $b\bar{b}\to hh$ (the rates change very little
when $m_h$ varies from 123 GeV to 127 GeV).

For each SUSY model we use the package NMSSMTools-3.2.0~\cite{NMSSMTools} to
scan over the parameter space and then select the samples which give a
SM-like Higgs boson in the range of $125 \pm 2 ~{\rm GeV} $ and also
satisfy various experimental constraints, including those listed in Section I.
The strategy of our scan is same as in \cite{July-Cao} except for three updates.
First, since the rare decay $B_s \to \mu^+ \mu^-$ has been recently observed with
$Br(B_s \to \mu^+ \mu^-) = 3.2^{+1.5}_{-1.2} \times 10^{-9}$~\cite{1213Bsmumu},
we use a double-sided limit
$0.8 \times 10^{-9} \leq  Br(B_s \to \mu^+ \mu^-) \leq 6.2 \times 10^{-9}$.
Second, for the LHC search of the non-standard Higgs boson,
we use the latest experimental data \cite{Htautau}.
The third one is that we require stops heavier than 200 GeV \cite{1213ThirdSquark-LHC}.
After the scan, we calculate the Higgs pair production rate in the allowed parameter
space. We will demonstrate the ratio $\sigma_{SUSY}/\sigma_{SM}$ for each surviving sample.
Of course, such a ratio is less sensitive to higher order QCD corrections.

\begin{figure}
\includegraphics[width=15cm]{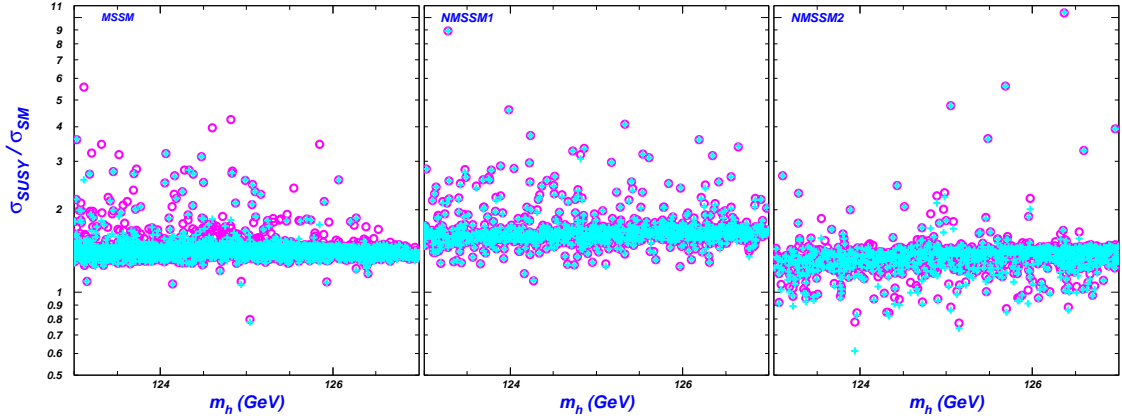}
\vspace{-0.5cm} \caption{The scatter plots of the surviving samples,
showing $\sigma_{SUSY}/\sigma_{SM}$ versus the SM-like Higgs boson mass.
The plus '+' (blue) denote the results with only the gluon fusion contribution,
while the circles '$\circ$' (pink) are for the total results. }
\label{fig-ratio-mh}
\end{figure}

In Fig.~\ref{fig-ratio-mh} we show the normalized production rate as a function of
the Higgs boson mass for the surviving samples in the MSSM and NMSSM
(for the NMSSM we show the results for the NMSSM1 and NMSSM2 scenarios defined in Sec.II).
This figure shows two common features for the three scenarios. One is that the production rate
can deviate significantly from the SM prediction: in most cases
the deviation exceeds $30\%$ and in some specail cases the  production rate
can be enhanced by one order.
The other feature is that for most cases the dominant contribution to the
pair production comes from the gluon fusion, which is reflected by the approximate
overlap of '$\circ$' (pink) with '+' (blue).
Fig.~\ref{fig-ratio-mh} also exhibits
some difference between different scenarios.
For example, in the MSSM the $b\bar{b}$ annihilation contribution can be dominant
for some surviving samples, which, however, never occurs in the NMSSM.
Another difference is that the NMSSM1 tends
to predict a larger production rate than other scenarios.

Now we explain some features of the results in Fig.~\ref{fig-ratio-mh}.
First, we investigate the cases of the MSSM where
the $b\bar{b}$ annihilation plays the dominant role in the production.
We find that they are characterized by a moderately large $\tan \beta$
($\tan \beta \sim 10$ so that the $Hb\bar{b}$ coupling is enhanced),
a moderately light $H$ ($ 300{\rm ~GeV} \lesssim m_H \lesssim 400{\rm ~GeV} $)
and a relatively large $Hhh$ coupling.
While for the NMSSM scenarios, since we are considering large $\lambda$ case,
only a relatively small $\tan \beta$ is allowed so that the $H_i b\bar{b}$
coupling is never enhanced sufficiently \cite{Feb-Cao}.
We also scrutinize the characters of the gluon fusion contribution in the MSSM.
As the first step, we compare the sbottom loop contribution with the stop loop.
We find that for the surviving samples the former is usually much smaller than the latter.
Next we divide the amplitude of Fig.~\ref{fig-diagram-total} into five parts
with $M_1,M_2,M_3,M_4$ and $M_5$ denoting the contributions from diagrams (1)+(2),
(3)+(4), (5), (6)+(7) and (8)+(9)+(10), respectively.
For each of the amplitude, it is UV finite so we can learn its relative size directly.
We find that the magnitudes of $M_2$ and $M_3$ are much larger than the others.
This can be understood
as follows: among the diagrams in Fig.~\ref{fig-diagram-total}, only (3), (4) and (5)
involve the chiral flipping of the internal stop, so in the limit $ m_{\tilde{t}_2},
m_{\tilde{t}_1} \gg 2 m_h$ the main parts of $M_2$ and $M_3$ can be written as
\begin{eqnarray}
M \sim \alpha_s^2 Y_t^2 ( c_1 \sin^2 2\theta_t \frac{A_t^2}{m_{\tilde{t}_1}^2} + c_2 \frac{A_t^2}{m_{\tilde{t}_2}^2}) \label{simpleform}
\end{eqnarray}
where $Y_t$ is the top quark Yukawa coupling,
$\theta_t$ and $A_t$ are  respectively the chiral mixing angle and the
trilinear soft breaking parameter in the stop sector, and $c_{1} $ and $c_2$
are ${\cal{O}}(1)$ coefficients with opposite signs.
Since a large $A_t$  is strongly favored to predict
$m_h \sim 125 {\rm ~GeV}$ in the MSSM \cite{Feb-Cao} and the other contributions are
usually proportional to $m_t^2/m_{\tilde{t}_i}^2$ or $m_h^2/m_{\tilde{t}_i}^2$, one can easily
conclude that $M_2$ and $M_3$ should be most important among the five amplitudes.
In fact, we checked that without the strong cancelation between
$M_2$ and $M_3$, the production rate can easily exceed 100 fb for most surviving samples.

\begin{figure}
\includegraphics[width=15cm]{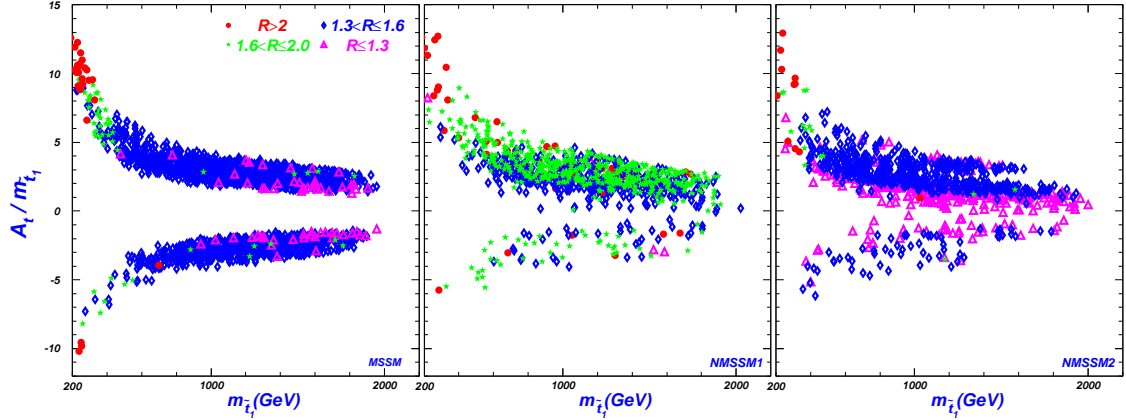}
\vspace{-0.5cm}
\caption{Same as Fig~\ref{fig-ratio-mh},
but showing $A_t/m_{\tilde{t}_1}$ versus $m_{\tilde{t}_1}$.
The samples are classified according to the value of
$R=\sigma_{SUSY}(gg\to hh)/\sigma_{SM}(gg\to hh)$
with $\sigma$ denoting the hadronic cross section via $gg\to hh$.}
\label{fig-ratio-atmst1mst1}
\end{figure}

As a proof for the validity of Eq.(\ref{simpleform}), in Fig.~\ref{fig-ratio-atmst1mst1} we
show $A_t/m_{\tilde{t}_1}$ versus $m_{\tilde{t}_1}$, where the samples
are classified according to the value of $R=\sigma_{SUSY}(gg\to hh)/\sigma_{SM}(gg\to hh)$.
The left panel indicates
that in the MSSM the region characterized by a light $m_{\tilde{t}_1}$ and a large $|A_t/m_{\tilde{t}_1}|$
usually predicts a large $R$. This can be understood as follows.
In the MSSM with a light  $\tilde{t}_1$,
the other stop ($\tilde{t}_2$) must be sufficiently heavy in order to predict
$m_h \sim 125 {\rm ~GeV}$ \cite{Feb-Cao}. Then, after expressing
$\sin^2 2 \theta_t$ in terms of $A_t$ and stop masses, one can find that the first term in
Eq.(\ref{simpleform}) scales like $ (A_t/m_{\tilde{t}_1})^4 (m_t^2 m_{\tilde{t}_1}^2/m_{\tilde{t}_2}^4)$,
and therefore its value grows rapidly with the increase of $|A_t/m_{\tilde{t}_1}|$ and is unlikely
to be canceled out by the second term in Eq.(\ref{simpleform}). In fact, the upper left
region of the panel reflects such a behavior. This panel also indicates that even for $\tilde{t}_1$
and $\tilde{t}_2$ at TeV scale, the production rate in the MSSM may still deviate
from its SM prediction by more than $30\%$. This is obvious since $|A_t|$ in Eq.(\ref{simpleform})
is usually larger than stop masses \cite{Feb-Cao}. Finally, we note
that for $m_{\tilde{t}_1} > 1 ~{\rm TeV}$, there exist some cases where the deviation
is small even for $A_t/m_{\tilde{t}_1} \sim 3$. We checked that these cases actually
correspond to a small mass splitting between $\tilde{t}_1$ and $\tilde{t}_2$.
In such a situation, the first term in
Eq.(\ref{simpleform}) is proportional to $A_t^2/m_{\tilde{t}_1}^2$ (since $\theta_t \simeq \pi/4$),
and its contribution to the rate is severely canceled by the second term.

\begin{figure}
\includegraphics[width=15cm]{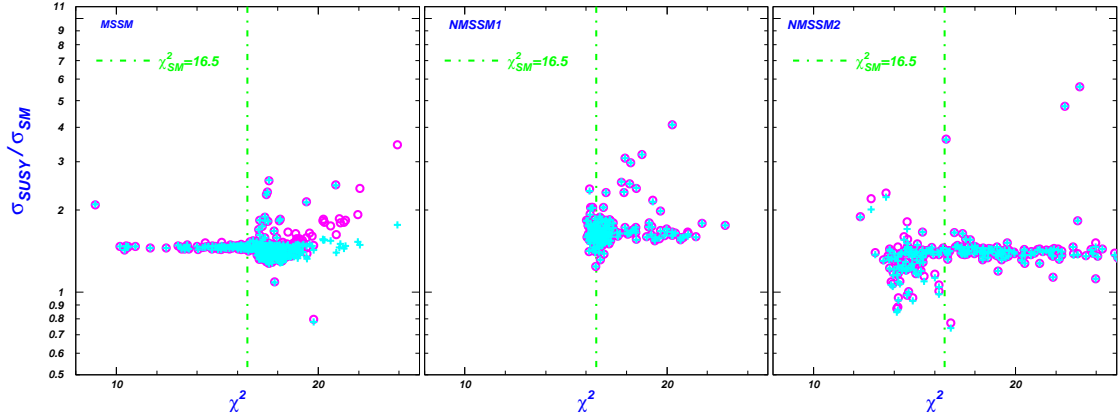}
\vspace{-0.5cm} \caption{Same as Fig.~\ref{fig-ratio-mh}, but
showing $\sigma_{SUSY}/\sigma_{SM}$ versus $\chi^2$.
Here only the samples satisfying 125 GeV $\le m_h \leq 126$ GeV are plotted.}
\label{fig-ratio-chisq}
\end{figure}

Eq.(\ref{simpleform}) may also be used to explain the results of the NMSSM1 scenario.
In this scenario we checked that the mixing effect on $m_h$ often exceeds the additional
tree level contribution (as discussed in Sec. II), and consequently the soft breaking
parameters in the stop sector are more tightly limited than the other two scenarios.
For example, given the same values of $m_{\tilde{t}_1}$ and $m_{\tilde{t}_2}$ for the three scenarios,
the NMSSM1 scenario usually prefers a larger $|A_t|$.
Consequently, this scenario tends to predict the largest production rate
according to Eq.(\ref{simpleform}).
As for the $R$ value in the NMSSM2 scenario, the situation is quite complex
because a large $\lambda$ alone can push the value of $m_h$ up to about 125 GeV
and thus the soft breaking parameters in the stop sector are not so constrained
by the Higgs mass \cite{Feb-Cao}. But, anyway, this scenario still
has the features that $R$ is maximized
for a large $A_t$ and a light $\tilde{t}_1$  and that $R$ can deviate
sizably from unity for TeV-scale stops.

Finally, we focus on the samples which predict a SM-like Higgs boson in the best fitted mass
region, $125 {\rm ~GeV} \le m_h \leq 126 {\rm ~GeV}$ \cite{GlobalFit}.
For these samples, we calculate the $\chi^2$ value with the LHC Higgs data
(for details, see \cite{July-Cao,GlobalFit}) and show its correlation with
the normalized rate $\sigma_{SUSY}/\sigma_{SM}$ in Fig.~\ref{fig-ratio-chisq}.
This figure indicates that in the MSSM and NMSSM2 scenarios, there exist a lot of samples
with $\chi^2$ much smaller than its SM value ($\chi^2_{SM} =16.5$),
which implies that the MSSM and NMSSM2 scenarios may be favored by the current
data \cite{July-Cao}. In contrast,
the NMSSM1 scenario can only slightly improve the fit.
From this figure we also see that
in the favored parameter space with a small $\chi^2$ the production rate
can sizably deviate from the SM prediction (
in the parameter space with a large $\chi^2$ the production rate can be
several times larger than the SM value).
For example, in the low $\chi^2$ region of the MSSM,
the normalized rate is approximately 1.45, while in
the NMSSM2 scenario the rate varies from 0.7 to 2.4.

\section{Summary and Conclusions}
\label{Sum}

Recently, the CMS and ATLAS collaborations announced the discovery of a new resonance
whose property is in rough agreement with the SM Higgs boson. But the nature of this
new state, especially its role in electroweak symmetry breaking, needs to be scrutinized.
So the most urgent task for the LHC is to test
the property of this Higgs-like boson by measuring all the possible
production and decay channels with high luminosity.
Among the production channels, the Higgs pair production is a rare process
at the LHC. Since it can play an important role for testing the Higgs self-couplings,
it will be measured at the LHC with high luminosity.

In this work we studied the pair production of the SM-like Higgs boson
in the popular SUSY models: the MSSM and NMSSM.
To make our study realistic, we first scanned the parameter space of
each model by considering various experimental constraints.
Then we examined the Higgs pair production in the allowed parameter space.
We found that for most cases in both models,
the dominant contribution to the pair production comes from the gluon fusion process
with its rate maximized at a moderately light $\tilde{t}_1$ and a large trilinear
soft breaking parameter $A_t$. The production rate can be sizably enhanced relative
to the SM prediction: $\sigma_{SUSY}/\sigma_{SM}$ can reach 10, and even for a TeV-scale
stop it can also exceed 1.3.
For each model we also calculated its $\chi^2$ with current Higgs data and
found that in the most favored parameter region the value of $\sigma_{SUSY}/\sigma_{SM}$
is approximately 1.45 in the MSSM, while in the NMSSM it varies from 0.7 to 2.4.

\section*{Acknowledgement}
We thank Jingya Zhu for helpful discussions. This work
was supported in part by the National Natural Science Foundation of
China (NNSFC) under grant No. 10775039, 11075045,
11275245, 11222548, 10821504, 11135003 and 11247268, and by the Project
of Knowledge Innovation Program (PKIP) of Chinese Academy of
Sciences under grant No. KJCX2.YW.W10.


\end{document}